\documentclass[journal]{vgtc}                     

\onlineid{0}

\vgtccategory{Research}

\vgtcpapertype{system}

\title{Py maidr: Bridging Visual and Non-Visual Data Experiences Through a Unified Python Framework}

\author{%
  \authororcid{JooYoung Seo}{0000-0002-4064-6012}, \authororcid{Saairam Venkatesh}{0009-0007-7810-2305}, \authororcid{Daksh Pokar}{0009-0005-9630-351X}, \authororcid{Sanchita Kamath}{0000-0001-6469-0360}, \authororcid{Krishna Anandan Ganesan}{0009-0007-7654-4548}
}

\authorfooter{
  \item JooYoung Seo, Saairam Venkatesh, Daksh Pokar, Sanchita Kamath, Krishna Anandan Ganesan are from University of Illinois Urbana-Champaign. E-mail: \{jseo1005\,$|$\,saairam2\,$|$\,dakshp2\,$|$\,ssk11\,$|$\,kag8\}@illinois.edu\,.
}

\abstract{%
Although recent efforts have developed accessible data visualization tools for blind and low-vision (BLV) users, most follow a ``design for them'' approach that creates an unintentional divide between sighted creators and BLV consumers. This unidirectional paradigm may perpetuate a power dynamic where sighted creators produce non-visual content boundaries for BLV consumers to access. This paper proposes a bidirectional approach, ``design for us,'' where both sighted and BLV collaborators can employ the same tool to create, interpret, and communicate data visualizations for each other. We introduce \texttt{Py maidr}, a Python package that seamlessly encodes multimodal (e.g., tactile, auditory, conversational) data representations into visual plots generated by \texttt{Matplotlib} and \texttt{Seaborn}. By simply importing \verb|maidr| package and invoking the \verb|maidr.show()| method, users can generate accessible plots with minimal changes to their existing codebase regardless of their visual dis/abilities. Our technical case studies demonstrate how this tool is scalable and can be integrated into interactive computing (e.g., Jupyter Notebook, Google Colab), reactive dashboards (e.g., Shiny, Streamlit), and reproducible literate programming (e.g., Quarto). Our performance benchmarks demonstrate that \texttt{Py maidr} introduces minimal and consistent overhead during the rendering and export of plots against \texttt{Matplotlib} and \texttt{Seaborn} vanilla baselines. This work significantly contributes to narrowing the accessibility gap in data visualization by providing a unified framework that fosters collaboration and communication between sighted and BLV individuals. Our live demo and example code are available at \url{https://xability.github.io/py-maidr}.

}


\teaser{
  \centering
  \includegraphics[width=0.85\columnwidth, alt={The interface is divided into four labeled sections. On the left, a blue-bordered panel titled "SHORTCUTS" lists keyboard commands for navigating and interacting with charts, such as arrow keys for movement and key combinations for toggling modes. In the top center, a red-bordered "SETTINGS" panel contains sliders, dropdowns, and text fields for customizing display, sonification parameters, and connecting to large language models like GPT, Claude, or Gemini. Below that, in an orange-bordered box labeled "CHART ASSISTANT," a chat window is shown with a message stating no agents are enabled. On the right, inside a purple-bordered panel labeled "BOX PLOT WITH BRAILLE AND TEXT MODES ON," a box plot displays life expectancy by continent. Below the plot is a text and Braille transcription of selected data: “Continent is Oceania, Life Expectancy, 0 Lower outlier(s),” followed by a Braille grid representing the data.}]{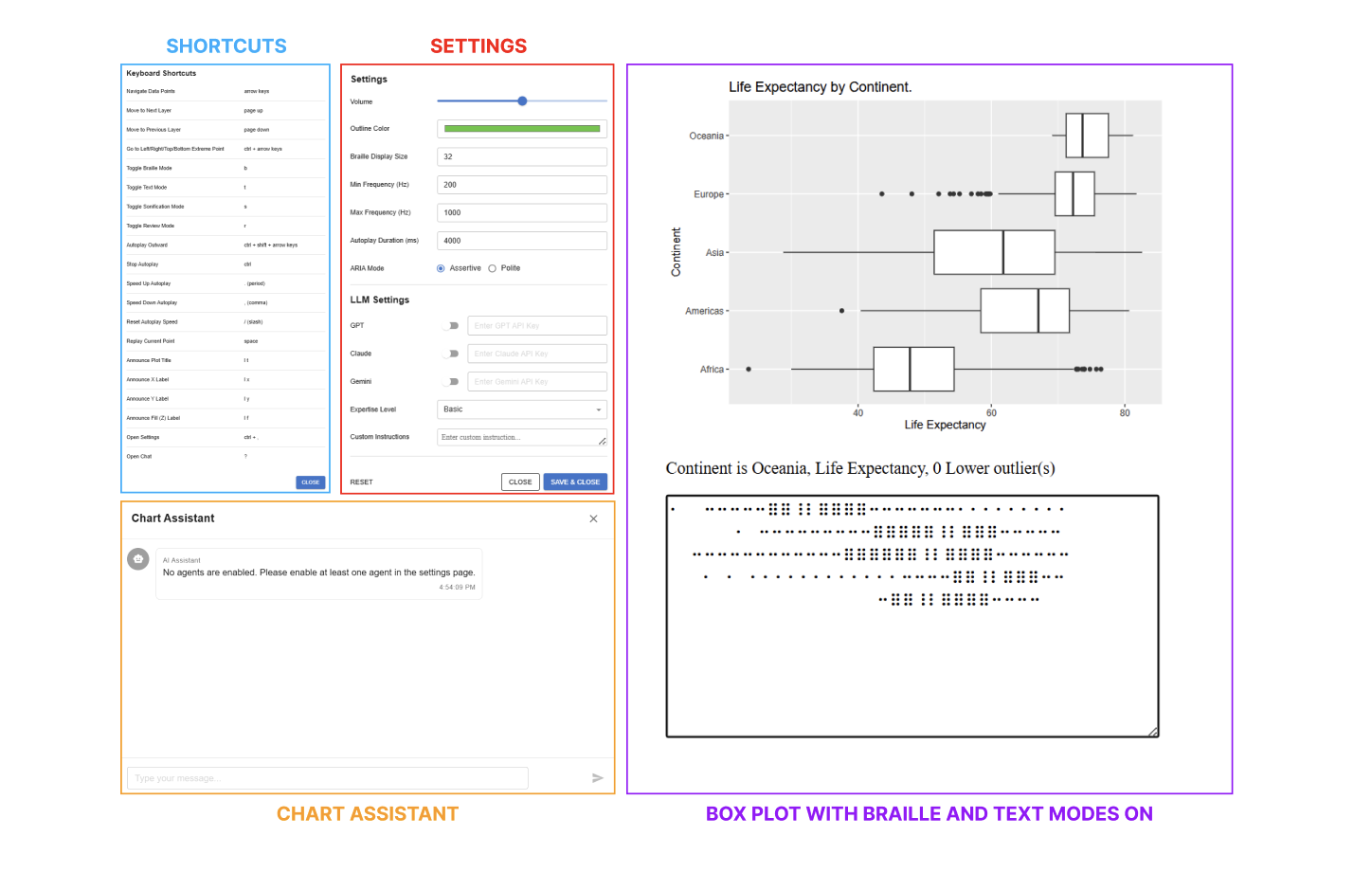}
  \caption{\texttt{Py maidr} Interface Panels 
  }
  \label{fig:teaser}
}

\renewcommand{\manuscriptnotetxt}{}

\graphicspath{{figs/}{figures/}{pictures/}{images/}{./}} 

\usepackage{tabu}                      
\usepackage{booktabs}                  
\usepackage{lipsum}                    
\usepackage{mwe}                       
\usepackage{minted}

\usepackage{xcolor}
\usepackage{listings}

\definecolor{myhighlight}{RGB}{255,255,150} 

\usepackage{mathptmx}                  

\begin{document}



\maketitle


\section{Introduction}
\label{sec:introduction}

Data visualization is fundamental to data science, enabling the discovery and communication of insights within complex datasets \cite{marriottInclusiveDataVisualization2021,kimAccessibleVisualizationDesign2021}. However, the reliance on purely visual representations often excludes individuals who are blind or have low vision (BLV) from fully participating in data analysis, interpretation, and collaborative workflows \cite{leeReachingBroaderAudiences2020,jooyoungseoTeachingVisualAccessibility2023}. Previous attempts addressed this issue with non-visual data representations, such as alt text and verbal descriptions \cite{lundgardAccessibleVisualizationNatural2022,morashGuidingNoviceWeb2015,aultEvaluationLongDescriptions2002,belleAltTexifyPipelineGenerate2022b}, data sonification \cite{summersConvertingGraphicalDatavisualizations2019,hoqueAccessibleDataRepresentation2023,AudioGraphsApple,julianna-langstonChart2Music2022,siuSupportingAccessibleData2022,kimErieDeclarativeGrammar2024}, audio-tactile graphs \cite{gotzelmannVisuallyAugmentedAudioTactile2018,heTacTILEPreliminaryToolchain2017,brownVizTouchAutomaticallyGenerated2012a}, and more recently with cross-modality approaches \cite{thompsonChartReaderAccessible2023,sharifVoxLens2022}. Furthermore, authoring tools and guidelines \cite{elavskyHowAccessibleMy2022,elavskyDataNavigatorAccessibilityCentered2023}, as well as workshop and training interventions \cite{sharifUnderstandingReducingChallenges2024,sharifWorkshopEducationalIntervention2024}, have been suggested to encourage creators of data visualizations to promote accessibility for a wide range of audiences with diverse abilities, including screen-reader users.
While these approaches have made strides in improving accessibility, most follow a ``design for them'' approach that creates an unintentional divide between sighted creators and BLV consumers. This unidirectional paradigm may perpetuate a power dynamic where sighted creators produce non-visual content \textit{for} BLV consumers to access. Furthermore, even when BLV users can independently create non-visual representations like sonifications or tactile graphs, their sighted collaborators may lack the familiarity or tools to interpret these modalities as fluently as they interpret visual charts. This creates a significant barrier to effective communication and equitable collaboration between sighted and BLV individuals.
This paper proposes a bidirectional approach, ``design for us,'' where both sighted and BLV collaborators can employ the same tool to create, interpret, and communicate data visualizations for each other. Extending our prior work on multimodal access and interactive data representation (MAIDR) framework \cite{seoMAIDRMakingStatistical2024,seoMAIDRMeetsAI2024}, we introduce \texttt{Py maidr}, a Python package that seamlessly encodes multimodal (e.g., tactile, auditory, conversational) data representations into visual plots generated by \texttt{Matplotlib} \cite{hunterMatplotlib2DGraphics2007} and \texttt{Seaborn} \cite{waskomSeabornStatisticalData2021}. By simply importing \verb|maidr| package and invoking \verb|maidr.show()| method for \texttt{Matplotlib} objects, users can automatically generate accessible plots with minimal changes to their existing codebase regardless of their visual dis/abilities.

\texttt{Py maidr} specifically targets Python and its dominant visualization libraries to leverage their widespread adoption. Python serves as the lingua franca for modern data science, machine learning, and scientific computing, consistently ranked as the top programming tool in developer surveys, data professional reports, and on open-source platforms \cite{staffOctoverseAILeads2024, 2022KaggleMachine, 2024StackOverflow}. Within this ecosystem, \texttt{Matplotlib} \cite{hunterMatplotlib2DGraphics2007} and \texttt{Seaborn} \cite{waskomSeabornStatisticalData2021} are the most widely adopted visualization libraries, used regularly by the vast majority of data scientists \cite{2022KaggleMachine}. Enhancing these ubiquitous tools offers the potential for broad accessibility impact across research, education, and industry.
Our technical case studies demonstrate \textit{Py maidr's} versatility and scalability, showcasing its integration into interactive computing environments (\textit{Jupyter Notebook}, \textit{Google Colab}), reactive web applications (\textit{Shiny}, \textit{Streamlit}), and reproducible literate programming documents (\textit{Quarto}). Performance benchmarks confirm that \texttt{Py maidr} introduces minimal and consistent computational overhead compared to baseline \texttt{Matplotlib} and \texttt{Seaborn} rendering.

\section{Background: Multimodal Access and Interactive Data Representation}
\label{sec:related_work}

Our prior work \cite{seoMAIDRMakingStatistical2024} proposed the Multimodal Access and Interactive Data Representation (MAIDR) design framework to enhance data visualization accessibility for screen-reader users. It features three togglable modalities (Braille, Text, Sonification - BTS), enabling up to eight representation combinations. Based on this framework, we released the \textit{maidr.js} open-source library \cite{XabilityMaidr2024} as a proof-of-concept. \textit{maidr.js} is a JavaScript library that takes a rendered SVG image and its declarative JSON-format metadata schema (e.g., plot type, data, CSS selectors for visual markers) similar to Vega-Lite \cite{satyanarayanVegaLiteGrammarInteractive2017} as input. It generates interactive and multimodal data representations (touchable Braille, readable Text, and audible Sonification) overlaid on the image. The maidr JSON schema plays a critical role in mapping non-visual modalities to SVG's visual elements with CSS selectors, allowing users to accurately trace each data point using arrow keys. Furthermore, this image-with-data approach helps mitigate hallucination effects when integrating AI-based visual question and answering (VQA) features into \textit{maidr.js} \cite{seoMAIDRMeetsAI2024}. However, since \textit{maidr.js} is a visual-agnostic front-end engine, creators must manually provide a rendered SVG and declare its corresponding JSON schema. This non-trivial workflow is impractical for most data visualization creators and time-consuming. To address this gap, we propose and release a Python binder for \textit{maidr.js} that automatically extracts and generates the maidr-compliant JSON schema from \texttt{Matplotlib} and \texttt{Seaborn} figures while rendering their associated SVG images. The following section focuses exclusively on our implementation of \texttt{Py maidr}, while details of the \textit{maidr.js} library can be found in our previous work \cite{seoMAIDRMakingStatistical2024,seoMAIDRMeetsAI2024}.

\section{Py maidr Architecture}
\label{sec:system_implementation}

The architecture of \texttt{Py maidr} adopts a three-tier structure comprising an interception layer, an extraction engine, and a rendering bridge, all tailored to address four core challenges that arise in imperative plotting workflows: \textbf{(1) Implicit Plot Semantics}: Visualization libraries like \texttt{Matplotlib} express visualizations procedurally through drawing commands rather than declarative descriptions. This lack of semantic metadata requires external reconstruction of chart structure and intent to be used in the core engine; \textbf{(2) API Heterogeneity}: Python visualization libraries often implement different API layers (e.g., scripting vs. object-oriented in \texttt{Matplotlib}), making unified metadata extraction non-trivial; \textbf{(3) State Management}: \texttt{Matplotlib} relies on a global state model that complicates plot disambiguation and makes it difficult to maintain referential integrity in multi-plot or interactive contexts; and \textbf{(4) Cross-Modal Synchronization}: For accessibility features like Braille and sonification to work seamlessly, visual elements must be precisely linked with their semantic representations in the DOM.

To tackle these challenges, \texttt{Py maidr} introduces a runtime architecture that captures plotting behavior through monkey patching, reconstructs plot semantics through polymorphic extraction logic, and injects rich metadata into the SVG output to support downstream interaction. The following sections detail the architecture and implementation of each component.

\begin{figure*}[ht]
    \centering
    \includegraphics[width=\textwidth, alt={The diagram is a left-to-right flowchart with four main vertical sections representing the system architecture of \texttt{Py maidr}. On the far left, a box labeled ``User Code'' points right to two stacked boxes labeled ``\texttt{Matplotlib}'' and ``\texttt{Seaborn}'' under the ``Plotting Libraries'' section. Arrows lead from both libraries into the next section, ``Interception Layer", which includes a large box labeled ``Monkey Patching'' and a smaller box below it labeled ``Context Manager". To the right, the ``Extraction Layer'' contains a sequence of three stacked boxes: ``Figure Manager", ``maidr Plot Factory", and ``maidr Plots'' with an additional box labeled ``Extractor Mixins'' branching from ``maidr Plots". This section extracts and structures plot semantics. The final section on the far right, ``Rendering Layer", has a box labeled ``SVG with maidr Data'' above a smaller ``Highlight Context'' box. These feed into the final output box at the bottom right labeled ``maidr Accessible Plot". }]{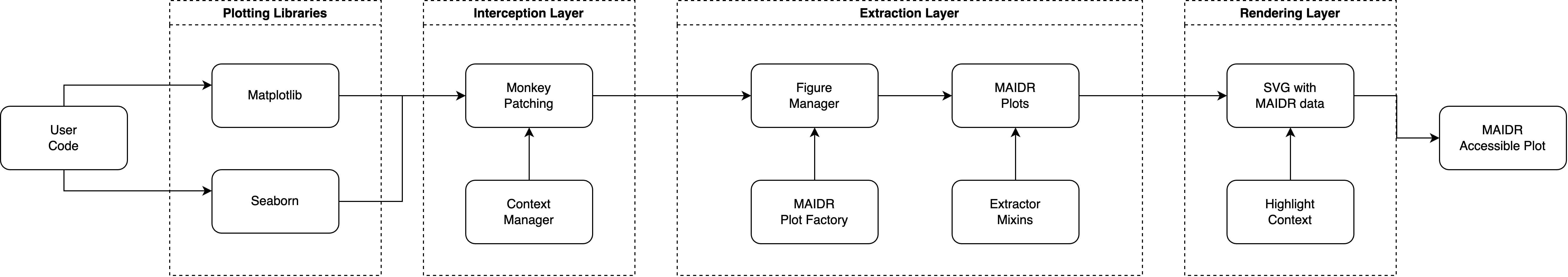}
    \caption{System architecture overview.}
    \label{fig:system-architecture}
\end{figure*}

\subsection{Layer 1: Plot Interception}
\label{subsec:pil}
This layer is responsible for seamlessly intercepting plot creation commands issued by users, without requiring any changes to their code. The interception mechanism is implemented through dynamic monkey-patching using the \texttt{wrapt} library, which allows the system to override plotting functions while preserving their original call signatures and behavior.

\subsubsection{Monkey-Patching Strategy}
\label{subsubsec:monkey-patching}
\texttt{Py maidr} targets a curated set of 87 functions spanning \texttt{Matplotlib}’s pyplot and Axes APIs, as well as \textit{Seaborn’s} high-level plotting functions. This includes both canonical function calls like \verb|plt.bar()| and context-specific methods like \verb|Axes.plot()| or \verb|sns.boxplot()|.
To ensure the patching is robust and non-invasive, the wrapped functions are decorated with a context-sensitive execution guard that prevents the system from responding to internal calls made by \texttt{Seaborn} or \texttt{Matplotlib} themselves. This prevents redundant or incorrect registration of plot data when, for example, \texttt{Seaborn} internally invokes \texttt{Matplotlib} primitives to render a single chart.

\subsubsection{Context Management}
\label{subsubsec:context-management}
\texttt{Py maidr} addresses internal call distinction using Python’s contextvars, which allow thread-local tracking of execution state. Each plotting call is evaluated in a context-sensitive manner, distinguishing between user-initiated and library-internal invocations.
A recursive call management strategy ensures that nested plotting calls (common in \texttt{Seaborn}) do not re-enter the extraction pipeline. This makes the interception layer stable across diverse usage patterns.

\subsubsection{Highlighting Integration}
\label{subsubsec:highlighting-integration}
During SVG rendering, each visual primitive that represents a data point—such as a bar, marker, or line segment—is tagged with a custom \textit{maidr} attribute to support downstream accessibility features. To accomplish this, we patch the drawing methods of \texttt{Matplotlib}’s core classes (namely, Patch, QuadMesh, Line2D, and PathCollection). Each time one of these primitives is drawn, we generate a globally unique identifier, assign it to the primitive via \texttt{Matplotlib}’s gid property, and record the mapping in an internal elements map.
When the figure is serialized to SVG through \texttt{Matplotlib}’s XMLWriter, we inject the unique identifier as a \textit{maidr} attribute on the corresponding SVG element. These attributes serve as anchors for an assistive front end, which can dynamically align user focus or navigation commands with the appropriate DOM elements. A dedicated HighlightContextManager governs this process, ensuring that only explicitly tracked elements receive annotations—thereby preserving rendering performance and preventing extraneous markup.

\subsection{Layer 2: Data Extraction Engine}
\label{subsec:dataextraction-engine}
The extraction engine translates raw plot objects into structured, accessible schemas. It operates on top of \texttt{Matplotlib}’s artist model and is implemented as a polymorphic hierarchy of extractor classes.

\subsubsection{Polymorphic Plot Processing}
\label{subsubsec:plot-processing}
The system defines an abstract base class (\verb|MaidrPlot|) that represents the semantic schema of a single plot. Concrete subclasses implement logic specific to various chart types—bar charts, line charts, scatter plots, heatmaps, and so on. Each subclass is responsible for identifying its relevant visual elements, extracting data values, and encoding axis semantics.
A factory pattern coordinates this polymorphism. When a plot creation is intercepted, the system determines the appropriate plot type and delegates instantiation to a factory that returns the correct handler. This design centralizes control logic while allowing chart-specific extraction to be encapsulated in modular units.
Each \verb|maidrPlot| instance includes information about its associated Axes, its position within a subplot grid (row and column index), and a set of DOM elements that should be tagged for highlighting. Subplots are a part of multi-view visualizations where multiple plots can be rendered at once, and each plot is referred to as subplot. The object also provides a method to output a declarative schema, which includes the chart’s type, axis labels, data points, and CSS selectors.

\subsubsection{Mixin-Based Specialization}
\label{subsubsec:mixin-specialization}
\texttt{Py maidr} introduces a set of functional mixins to address recurring sub-tasks within plot extraction logic. These include: \textbf{(1) Container Extraction}: Handles nested artist containers like BarContainer, which encapsulate multiple patches representing grouped or stacked bars; \textbf{(2) Level Extraction}: Resolves categorical axis levels, especially when tick labels are manually set or when dealing with \texttt{Seaborn}’s abstracted interfaces; and \textbf{(3) Dictionary Merging}: Merges independently computed metadata fragments (e.g., axes info and data values) into a single cohesive schema.
This design encourages reuse, improves maintainability, and avoids duplicating boilerplate logic across multiple plot types.

\begin{table*}[ht]
  \centering
  \caption{Performance Evaluation of \texttt{Py maidr} Overhead (in milliseconds)}
  \label{tab:performance}
  \begin{scriptsize}
  \begin{tabular*}{\textwidth}{@{\extracolsep{\fill}} lccccccc }
    \toprule
    \textbf{Plot Type} 
      & \multicolumn{3}{c}{\textbf{Matplotlib}} 
      & \multicolumn{3}{c}{\textbf{Seaborn}} \\
    \cmidrule(lr){2-4}\cmidrule(lr){5-7}
      & \textbf{With \texttt{Py maidr}} & \textbf{Without \texttt{Py maidr}} & \textbf{Overhead $\Delta$} 
      & \textbf{With \texttt{Py maidr}} & \textbf{Without \texttt{Py maidr}} & \textbf{Overhead $\Delta$} \\
    \midrule
    Bar             & $41.5\pm8.3$    & $40.3\pm9.2$    & $+1.2$  
                    & $49.9\pm9.6$    & $47.5\pm12.2$  & $+2.4$  \\
    Horizontal Box  & $38.9\pm11.9$   & $38.3\pm10.7$   & $+0.6$  
                    & $657\pm20$      & $654\pm19$     & $+3.0$  \\
    Vertical Box    & $40.6\pm12.1$   & $39.8\pm11.2$   & $+0.8$  
                    & $658\pm20$      & $653\pm22$     & $+5.0$  \\
    Line            & $33.9\pm4.7$    & $32.8\pm4.7$    & $+1.1$  
                    & $135\pm13$      & $135\pm15$     & $0.0$   \\
    Dodged          & $15.5\pm6.9$    & $14.9\pm4.9$    & $+0.6$  
                    & $276\pm15$      & $272\pm19$     & $+4.0$  \\
    Multilayered    & $42.4\pm14.4$   & $41.4\pm19.9$   & $+1.0$  
                    & $245\pm16$      & $238\pm27$     & $+7.0$  \\
    Multipanel      & $85.1\pm20.7$   & $83.1\pm34.8$   & $+2.0$  
                    & $178\pm21$      & $171\pm40$     & $+7.0$  \\
    Scatter         & $19.2\pm4.7$    & $17.8\pm4.6$    & $+1.4$  
                    & $41.6\pm2.4$    & $39.3\pm2.8$   & $+2.3$  \\
    Histogram       & $99.1\pm39.8$   & $94.2\pm11.7$   & $+4.9$  
                    & $254\pm83$      & $253\pm10$     & $+1.0$  \\
    Stacked         & $24.1\pm6.4$    & $22.4\pm5.6$    & $+1.7$  
                    & $73\pm8.8$      & $71.6\pm5.4$   & $+1.4$  \\
    Heatmap         & $41.6\pm1.7$    & $40.4\pm3.4$    & $+1.2$  
                    & $171\pm13$      & $167\pm17$     & $+4.0$  \\
    Multiline       & $21.5\pm1.5$    & $20.3\pm6.4$    & $+1.2$  
                    & $431\pm21$      & $429\pm23$     & $+2.0$  \\
    \midrule
    \textbf{Overall} 
                    & $41.95\pm11.09$ & $40.47\pm10.59$ & $1.48\pm1.15$ 
                    & $264.13\pm20.23$& $260.87\pm17.70$& $3.26\pm2.23$ \\
    \bottomrule
  \end{tabular*}
  \end{scriptsize}
\end{table*}

\subsection{Layer 3: Rendering Bridge}
\label{subsec:rendering-bridge}
The final layer of the system is responsible for embedding the extracted plot metadata into the output SVG and HTML, making it accessible to client-side assistive technologies.

\subsubsection{SVG Metadata Injection}
\label{subsubsec:metadata-injection}
During rendering, the system exports the \texttt{Matplotlib} figure to SVG and post-processes the output to inject two forms of metadata: \textbf{(1) Structural Attributes}: Each visual element (e.g., a bar or point) is tagged with \verb|maidr="true"|, allowing CSS and JavaScript selectors to reference them unambiguously; \textbf{(2) Semantic Payload}: The complete plot schema—extracted by \verb|MaidrPlot| objects—is serialized to JSON and embedded in a \verb|maidr-data| attribute at the root of the SVG. This allows the frontend to reconstruct the full semantic structure of the visualization without needing to parse or interpret the raw SVG.
These dual annotations enable precise coordination between visual, textual, and sonified representations.

\subsubsection{Environment-Adaptive Delivery}
\label{subsubsec:env-delivery}
\textit{Py maidr }supports multiple delivery targets. In interactive environments like Jupyter notebooks or VS Code, the output is rendered inside an iframe that resizes dynamically based on its content. For standalone use, the rendered HTML can be saved to a temporary file and launched in the system browser. In Shiny applications, the system integrates with reactive widgets to support live updates.
Each output mode preserves full accessibility support, including keyboard navigation and toggleable Braille/text interfaces.

\subsection{End-to-End Workflow}
From the user’s perspective, the experience is seamless and requires no changes to their plotting code. The internal workflow proceeds as follows: \textbf{(1) User Executes Plotting Code}: A call is made to a plotting function, such as \verb|sns.barplot()| or \verb|plt.scatter()|; \textbf{(2) Intercept and Classify}: The function is intercepted via a patch. \texttt{Py maidr} determines the plot type (e.g., BAR, LINE, STACKED) based on input arguments and inferred semantics; \textbf{(3) Context Validation}: The system checks whether the call is internal (originating from \texttt{Seaborn} or \texttt{Matplotlib} itself) and suppresses processing if so. This avoids extracting metadata from helper plots created during high-level plotting; \textbf{(4) Plot Object Registration}: Once the true user-generated plot is identified, the FigureManager registers the plot under the current Figure. The appropriate \verb|MaidrPlot| handler is instantiated and populated; \textbf{(5) Rendering Trigger}: When the user calls \verb|maidr.show()| or \verb|maidr.save_html()|, the system renders the Figure to SVG, injects metadata, and generates a complete HTML document; and \textbf{(6) Frontend Activation}: A lightweight  JavaScript/TypeScript engine loads the \verb|maidr-data| schema and binds each data point to DOM elements, enabling multimodal exploration.    

\subsection{Performance Benchmark Evaluation}
\label{sec:eval}
To evaluate the performance impact of \textit{Py maidr}’s data extraction layer, we benchmarked the time to render and save statistical plots as SVGs under two conditions: \textbf{(1) Without Import}, using native \texttt{Matplotlib}/\texttt{Seaborn} rendering; and \textbf{(2) With Import}, using monkey-patched functions from \texttt{Py maidr} that inject accessibility metadata.

Despite the large data volume, \texttt{Py maidr} introduced only minimal overhead—on average $1.48\pm1.15$ ms for \texttt{Matplotlib} and $3.26\pm2.23$ ms for \texttt{Seaborn}. These differences represent relative increases of about 3.7\% and 1.3\%, respectively. The overhead remained consistent across plot types and stems from the semantic extraction and SVG augmentation required for multimodal accessibility (\cref{tab:performance}).

\section{Use Cases}
\label{sec:usecase}

\subsection{General Usage}


The following code snippet demonstrates how to use \texttt{Py maidr} in conjunction with \texttt{matplotlib}. Once rendered, an interactive plot is displayed in the browser, and users can either click or tab to activate the plot where they can navigate through data points with arrow keys. Users can toggle on and off and mix each modality via BTS keys\footnote{\texttt{Py maidr} supports the following plot types: simple barplot, stacked barplot, dodged barplot, histogram, single and multi lineplot, horizontal and vertical boxplot, heatmap, scatterplot, multi-layered plot, multi-panel plot, subplot, and faceted plot. All the live examples are available at \url{https://py.maidr.ai/examples}.}.







\begin{lstlisting}
(*@\colorbox{myhighlight}{ \# Install the latest maidr version}@*)
(*@\colorbox{myhighlight}{ !pip install -U maidr}@*)
import matplotlib.pyplot as plt
import numpy as np
(*@\colorbox{myhighlight}{ \# Just add this line to import `maidr'}@*)
(*@\colorbox{myhighlight}{ import maidr}@*)
(*@\colorbox{myhighlight}{ \# Assign a figure to a variable}@*)
(*@\colorbox{myhighlight}{ fig = plt.plot(np.arange(10), np.arange(10)**2)}@*)
plt.xlabel('x')
plt.ylabel('y')
# plt.show()
(*@\colorbox{myhighlight}{ \# Display figure like `plt.show()'}@*)
(*@\colorbox{myhighlight}{ maidr.show(fig)}@*)
(*@\fcolorbox{myhighlight}{myhighlight}{\parbox[t]{\dimexpr\linewidth-1\fboxsep-1\fboxrule}{
\# Uncomment the following line to save and share the accessible version of your plot with others!
}}@*)
# maidr.save_html(fig, ``output.html'' )
\end{lstlisting}

\subsection{Interactive Computing Environments}

Interactive computing environments like Jupyter Notebook, JupyterLab, and Google Colab support collaborative data analysis; however, BLV users often encounter barriers when accessing visualizations \cite{potluriNotablyInaccessibleData2023}. \texttt{Py maidr} integrates seamlessly with these platforms, embedding interactive figures in iframe sandboxes to prevent JavaScript interference and key conflicts. Examples and live demos are available at \url{https://py.maidr.ai/examples.html#interactive-\\ computing-jupyter-notebooks-jupyter-labs-google-colab}.

\subsection{Reactive Web Dashboards}

Reactive web dashboards have long been inaccessible to BLV users, often requiring a specialized screen reader–friendly version \cite{srinivasanAzimuthDesigningAccessible2023}. With \texttt{Py maidr}, reactive dashboards and visualizations can be made accessible to both sighted and BLV users, as it supports two popular web frameworks: \texttt{Shiny} and \texttt{Streamlit}. The codebase and working examples are available at \url{https://py.maidr.ai/examples.html#reactive-dashboard}.








\subsection{Reproducible Literate Programming Documents}

\texttt{Py maidr} is designed to work with Quarto, a scientific and technical publishing system \cite{allaireQuartoOpensourceScientific2022}. Quarto enables the creation of dynamic documents, reports, and presentations using Pandoc-flavored Markdown syntax \cite{macfarlanePandocUniversalMarkup2023}. By integrating \texttt{Py maidr} with Quarto, users can produce reproducible visualizations accessible to both sighted and BLV users. The \texttt{Py maidr} user manual is also written in Quarto, and its codebase is available at \url{https://github.com/xability/py-maidr/tree/main/docs}.

\section{Conclusion}
\label{sec:conclusion}

Data visualization is vital to data science but remains largely inaccessible to BLV individuals. \texttt{Py maidr} addresses this gap through a ``design for us'' approach, embedding multimodal outputs (Braille, text, sonification) into familiar plotting libraries like \texttt{Matplotlib} and \texttt{Seaborn}. It requires minimal integration and works across Jupyter, web dashboards, and Quarto documents with negligible overhead. Future work includes expanding to \texttt{Plotly} and developing \texttt{R maidr} for \texttt{ggplot2}. Our ultimate vision is a future where data visualization is inherently accessible, empowering everyone to participate fully in the data-driven world.

\bibliographystyle{abbrv-doi-hyperref}

\bibliography{references/a11y_viz}

\appendix 







\end{document}